\documentclass[12pt]{article}
\usepackage{amsmath,amssymb}
\usepackage{graphicx}

\begin{document}

\thispagestyle{empty}
\begin{flushright}
OU-HET 682
\\
\end{flushright}
\vskip3cm
\begin{center}
{\Large {\bf Multiple D4-D2-D0 on the Conifold
 and
\\[0.3cm]
Wall-crossing with the Flop}}
\vskip1.5cm
{\large 
{Takahiro Nishinaka\footnote{nishinaka [at] het.phys.sci.osaka-u.ac.jp}
}
}
\vskip.5cm
{\it Department of Physics, Graduate School of Science, 
\\
Osaka University, Toyonaka, Osaka 560-0043, Japan}
\end{center}

\vskip1cm
\begin{abstract}
 We study the wall-crossing phenomena of D4-D2-D0 bound states with two units of D4-brane charge on the resolved conifold. We identify the walls of marginal stability and evaluate the discrete changes of the BPS indices by using the Kontsevich-Soibelman wall-crossing formula. In particular, we find that the field theories on D4-branes in two large radius limits are properly connected by the wall-crossings involving the flop transition of the conifold. We also find that in one of the large radius limits there are stable bound states of two D4-D2-D0 fragments.
\end{abstract}


\newpage

\setcounter{tocdepth}{2}
\tableofcontents

\section{Introduction}

BPS states in theories with extended supersymmetry have attracted much attention in the study of non-perturbative phenomena. These states are stable against decay in most cases because they belong to the short multiplet of supersymmetry. Accordingly their degeneracy (or index) is piecewise constant in the moduli space, but it discretely changes when the moduli cross the ``walls of marginal stability.'' Since these walls are real codimension one subspace, the moduli space can be divided into chambers surrounded by marginal stability walls, and the index is exactly constant in each chamber.

Recently, there has been remarkable progress in the study of these wall-crossing phenomena, especially in string theory on a Calabi-Yau three-fold. In the small string coupling regime, the BPS states are described by wrapped D-branes on supersymmetric cycles in the Calabi-Yau manifold. Their wall-crossing is associated with the appearance or disappearance of D-brane bound states in the spectrum. When the Calabi-Yau moduli cross the walls, some D-brane bound states cease to exist or newly appear in the spectrum. The wall-crossing of the BPS D-branes were studied from various point of view \cite{Szendroi, Diaconescu-Moore, Jafferis-Saulina, Cheng-Verlinde, Andriyash-Moore, NN, Nagao, CDWM, Collinucci-Wyder, Jafferis-Moore, Chuang-Jafferis, Ooguri-Yamazaki1, CSU, Dimofte-Gukov, David, Manschot1, Chuang-Pan, AOVY, Herck-Wyder, Nagao-Yamazaki, Sulkowski, Lee-Yi, Aganagic-Yamazaki, DGS, Szabo:2009vw, Krefl, CDP1, Manschot2, CDP2, Ooguri:2010yk, Aganagic-Schaeffer, Nishinaka:2010qk}. On the other hand, from the four-dimensional supergravity point of view, these wall-crossing phenomena are related to the existence of multi-centered BPS solutions. How many multi-centered solution exist in the spectrum depends on the boundary conditions of the Calabi-Yau moduli field at spatial infinity \cite{Denef:2000nb, Denef:2002ru, Denef-Moore, AdS3-S2, BEMB}.

One of the most remarkable progress is the work by Kontsevich and Soibelman\cite{Kontsevich-Soibelman} (see also \cite{Kontsevich-Soibelman2}). They proposed a mathematical wall-crossing formula which tells us how the BPS index changes at the walls of marginal stability.\footnote{Primitive and semi-primitive wall-crossing formulae were already proposed in \cite{Denef-Moore} by supergravity analysis.} By using this formula, we can learn the BPS degeneracy in various chambers in the moduli space. The physical meaning of this formula was studied in \cite{GMN1, GMN2, Cecotti-Vafa, GMN3, Andriyash:2010qv, Andriyash:2010yf}.

In this paper, we study the wall-crossing phenomena of D4-D2-D0 bound states with {\em two units} of D4-brane charge on the resolved conifold. The resolved conifold is a non-compact Calabi-Yau three-fold which has one compact 2-cycle and no compact 4-cycle. We introduce two D4-branes on the same non-compact supersymmetric 4-cycle and evaluate the BPS index of the D2-D0 bound states on the D4-branes. By changing the K\"ahler moduli of the conifold, which are the size and the B-field for the compact two-cycle, various walls of marginal stability are crossed.
Using the Kontsevich-Soibelman formula, we study the jumps in the BPS indices at the walls of marginal stability.
In particular, we find that the field theories on D4-branes in two large radius limits are consistently connected by wall-crossings, even when we have two units of D4-brane charge. These two large radius limits belong to topologically different resolutions of the conifold, which are connected by a topology-changing process called ``flop transition.'' (see for example \cite{Aspinwall:2004jr}) Therefore, our result shows that even in the case of two D4-branes the wall-crossing formula is compatible with the flop transition of the conifold.

In fact, this paper is a generalization of the work of \cite{Nishinaka:2010qk} where the wall-crossing of D4-D2-D0 bound states with {\em one} unit of D4-brane charge was studied. The main difference from \cite{Nishinaka:2010qk} is that in two D4-branes case we have to take into account the primitive wall-crossings in addition to the semi-primitive wall-crossings. Then the chamber structure in the moduli space is rather complicated and it turns out to be a hard task to evaluate the discrete change of the BPS index for each wall-crossing. Nevertheless, we can easily compare the BPS indices in two large radius limits where the field theory description on the D4-branes becomes reliable. The result shows that even in the large radius limit there exist the BPS bound states of two D4-D2-D0 fragments. This observation is similar to that in the work of \cite{Manschot1, Manschot2}. Furthermore, in the case of two D4-branes it turns out to be very natural to use the rational invariants $\overline{\Omega}(\Gamma)$ obtained from ordinary integer BPS indices $\Omega(\Gamma)$ by
\begin{eqnarray}
 \overline{\Omega}(\Gamma) &=& \sum_{m|\Gamma}\frac{\Omega(\Gamma/m)}{m^2}.
\end{eqnarray}
This was used in \cite{Manschot2} to obtain the S-duality invariant generating function.

This paper is organized as follows. In Section 2 we briefly review one D4-brane case in order to fix our notation. In Section 3 we study the case of two D4-branes. We introduce two non-compact D4-branes wrapped on the same supersymmetric divisor and study the D2-D0 bound states on them. There are two kinds of marginal stability walls, one for the primitive wall-crossings and the other for the semi-primitive wall-crossings. We study the discrete change in BPS indices at these walls and find that the field theories on D4-branes in two large radius limits are consistently connected by wall-crossings.
Section 4 contains a summary with some discussions. We also have an appendix with useful details.

\section{Review of one D4-brane case}

We here fix our notation and briefly review the result of the previous paper \cite{Nishinaka:2010qk}. The main topic in \cite{Nishinaka:2010qk} are wall-crossing of BPS bound states on the resolved conifold that consist of {\em one} D4-brane and arbitrary numbers of D2 and D0-branes. The resolved conifold $\mathcal{O}(-1)\oplus\mathcal{O}(-1) \to \mathbb{P}^1$ has only one compact 2-cycle $\mathbb{P}^1$ and no compact 4-cycle in it. We put a D4-brane on a non-compact supersymmetric divisor $\mathcal{O}(-1)\to \mathbb{P}^1$ in the conifold. However, if we consider the flop transition of the conifold, the topology of the 4-cycle is changed. After the flop, the D4-brane is wrapped on the whole fiber directions and localized on the rigid $\mathbb{P}^1$.

The electric and magnetic charges of the BPS bound states of interest can be written in terms of even-forms on the conifold $X$ as
\begin{eqnarray}
 \gamma = \mathcal{D} + k\beta - ldV,
\end{eqnarray}
where $\mathcal{D}\in H^{2}(X,\mathbb{R}),\,\beta \in H^4(X,\mathbb{R})$ and $dV\in H^6(X,\mathbb{R})$. The integers $k$ and $l$ denote D2 and D0-brane charges, respectively. We also note that $\beta$ is dual to the compact 2-cycle and therefore we only consider compact D2-branes. The 2-form $\mathcal{D}$ of course represents one unit of the non-compact D4-brane charge. Since our D4-brane is first on $\mathcal{O}(-1)\to \mathbb{P}^1$, the intersection between $\mathcal{D}$ and $\beta$ is
$
 \int \mathcal{D}\wedge \beta = -1.
$
The intersection product of charges are defined as
\begin{eqnarray}
 \left<\gamma_1,\gamma_2\right> &=& \int_X \gamma_1\wedge \check{\gamma}_2,
\end{eqnarray}
where $\check{\gamma}$ is an even-form obtained from $\gamma$ by inverting the sign of the 2-form and 6-form.

We can write the complexified K\"ahler parameter of the conifold as
\begin{eqnarray}
 t = z\mathcal{P} + \Lambda e^{i\varphi}\mathcal{P}',
\end{eqnarray}
where $z$ denotes the K\"ahler parameter for the rigid $\mathbb{P}^1$. The second term denotes the K\"ahler parameter for other non-compact cycles, and the local limit $\Lambda\to +\infty$ should be taken in the final result \cite{Jafferis-Moore}. The moduli space of interest is therefore the complex one-dimensional space of $z$ which denotes the size and the B-field for the compact two-cycle. The phase $\varphi$ is related to the ratio of the real and imaginary part of the K\"ahler parameter for non-compact cycles, which we take so that $\pi/4<\varphi<\pi/2$. Since the 4-form $\beta$ is dual to the rigid $\mathbb{P}^1$, we use the normalization as follows:
\begin{eqnarray}
 \int \mathcal{P}\wedge \beta = 1,\quad \int \mathcal{P}'\wedge \beta = 0. \label{eq:intersection}
\end{eqnarray}
In the moduli region ${\rm Im}\,z >0$, the 4-form $\beta$ represents the D2-brane charge in the K\"ahler cone, while in the region ${\rm Im}\,z <0$ it denotes one unit of {\em anti} D2-brane charge. The flop transition of the conifold relates these two regions in the moduli space. 

\begin{figure}
\begin{center}
 \includegraphics[height=3.5cm]{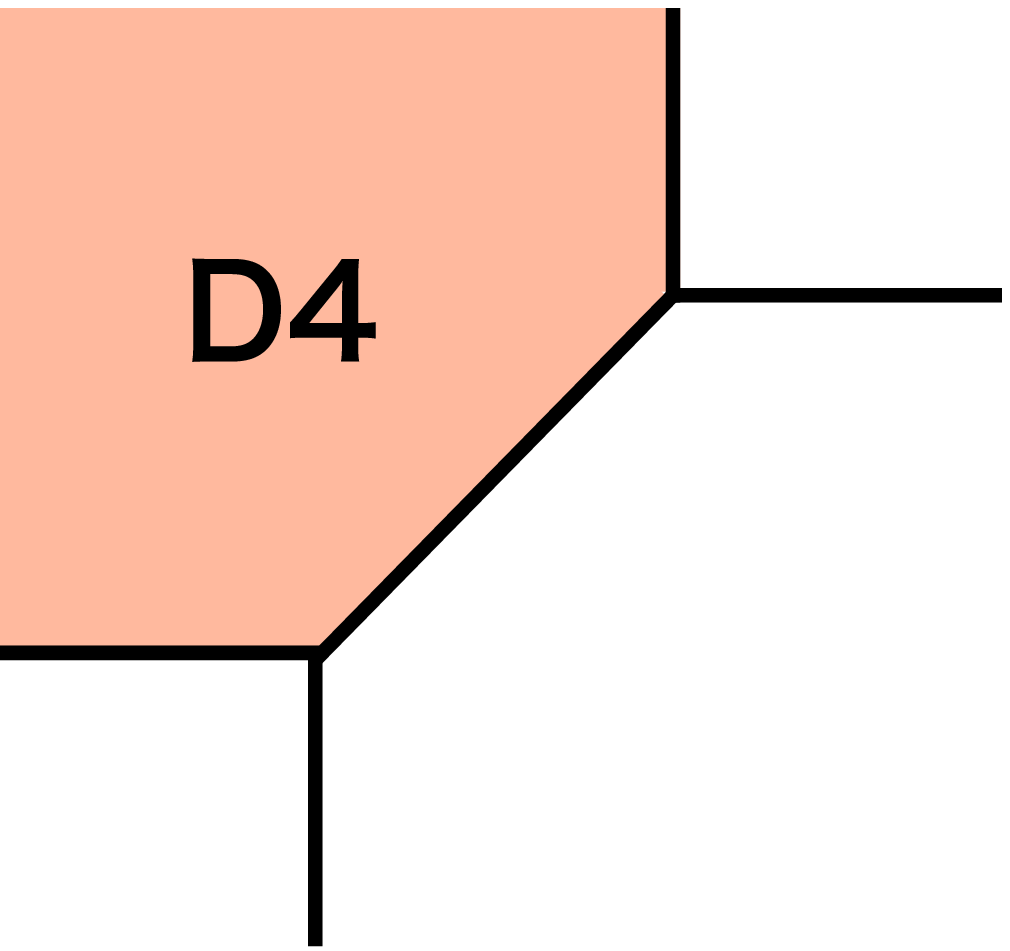}
 \hspace{2cm}
 \includegraphics[height=3.5cm]{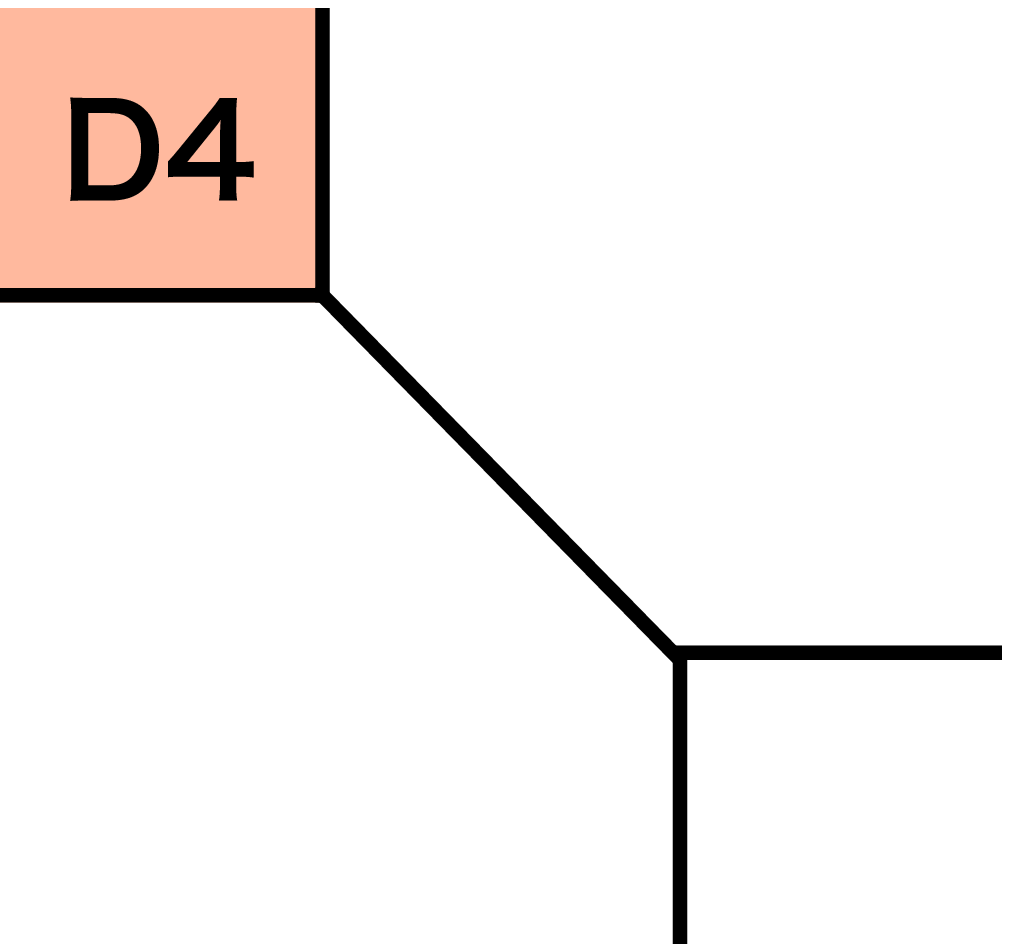}
\caption{Toric web diagrams of the two different resolutions of the conifolds. Our D4-branes are wrapped on $\mathcal{O}(-1)\to\mathbb{P}^1$ in the left picture, while in the right one it is wrapped on the whole fiber directions and localized on $\mathbb{P}^1$. These two resolutions are connected by the flop transition.}
\label{fig:flop}
\end{center}
\end{figure}

In the language of toric web diagrams, the flop transition can be shown as in Fig.~\ref{fig:flop}.
Suppose that we move the K\"ahler parameter $z$ from ${\rm Im}\,z = +\infty$ to ${\rm Im}\,z = -\infty$. When ${\rm Im}\,z=0$, the size of the compact two-cycle becomes zero and the topology of the conifold changes. However, no singularity occurs in the theory here if we fix ${\rm Re}\,z\not\in \mathbb{Z}$. Two moduli regions of ${\rm Im}\,z >0$ and ${\rm Im}\,z<0$ correspond to the left and right pictures in Fig \ref{fig:flop}, respectively. Two large radius limits previously mentioned are ${\rm Im}\,z = \pm \infty$, where the field theory description on the D4-branes becomes reliable. 

The central charge of a BPS state with the electromagnetic charge $\gamma$ is given by
$
 Z_{\rm norm}(\gamma) = \left<\gamma, {\bf \Omega}\right>
$, where ${\bf \Omega}$ is the normalized period vector.  Since the resolved conifold is non-compact, ${\bf \Omega} = -e^{t}$ follows up to a real positive prefactor.\footnote{This prefactor is now irrelevant because, as is shown below, we only need to evaluate the phase of the central charge in order to identify the walls of marginal stability.} So we can evaluate the holomorphic central charge as
\begin{eqnarray}
Z(\gamma) = \left<\gamma, -e^{t}\right> = -\int_{X}\gamma\wedge e^{-t}.\label{eq:central-charge}
\end{eqnarray}
This expression of the central charge is used in order to identify the walls of marginal stability in the moduli space.

The wall of marginal stability is defined as a codimension one subspace in the moduli space where the BPS states can decay into other BPS states. Such a decay can occur only if all the BPS states involved in the decay keep the same supersymmetry, namely, only if their central charges have the same phases. In \cite{Nishinaka:2010qk} it was found that the only possible decay channels for the BPS states with charge $\gamma = \mathcal{D} + k\beta -ldV$ are of the form\footnote{In fact, there is another type of possible decay channels of the form $\gamma \to (\gamma-\gamma_2) + (\gamma_2=-ndV)$. In \cite{Nishinaka:2010qk} it was shown, however, that the BPS index has no jump at the walls of marginal stability associated with this type of decay channels since we have no D6-brane.}
\begin{eqnarray}
 \gamma=\mathcal{D}+k\beta-ldV \;\;\to\;\; (\gamma_1 = \gamma-\gamma_2) \;+\; (\gamma_2 = \pm\beta-ndV).
\end{eqnarray}
We denote these walls as $\{W_{n}^{\pm}\}$. The D2-brane charge of $\gamma_2$ must be $\pm 1$ because the only non-vanishing index of D2-D0 bound states on the conifold are known to be \cite{Gopakumar-Vafa1, Gopakumar-Vafa2}
\begin{eqnarray}
 \Omega(\pm \beta + ndV) = 1.\label{eq:D2-D0}
\end{eqnarray}
The locations of the walls of marginal stability in the moduli space are identified by solving the equation $\arg[Z(\gamma)] = \arg[Z(\gamma_2)]$. 
By recalling Eqs. \eqref{eq:intersection} and \eqref{eq:central-charge}, we can evaluate the central charges of D4-D2-D0 and D2-D0 states as
\begin{eqnarray}
 Z(\gamma) \sim -\frac{c_4}{2}\Lambda^2e^{2i\varphi},\quad Z(\gamma_2) = m\beta + n,
\end{eqnarray}
respectively, in the local limit $\Lambda\to+\infty$. Here $c_4=\int \mathcal{D}\wedge \mathcal{P}'\wedge \mathcal{P}'$ and we assume $c_4>0$ without loss of generality. By solving $\arg[Z(\gamma)] = \arg[Z(\gamma_2)]$, the walls of marginal stability for BPS states with $\gamma = \mathcal{D} + l\beta -kdV$ are identified as in Fig. 2 in \cite{Nishinaka:2010qk}.

\subsection{Wall-crossing with topology-changing process}

One of the most important fact revealed in \cite{Nishinaka:2010qk} is that the D4-brane world-volume theories in the two different large volume limits can be related to each other via wall-crossing.

In the large volume limit where the sizes of all supersymmetric cycles become large, the physics of BPS wrapped D-branes are described by their world-volume theory. In particular, the BPS index (or BPS partition function) can be evaluated in the field theory on D-branes. Since we now have only one compact supersymmetric cycle $\mathbb{P}^1$, the large volume limit is the large radius limit of the rigid $\mathbb{P}^1$. But in our setup there are two such limits, namely ${\rm Im}\,z \to \pm\infty$. These two limits belong to the topologically-different resolutions of the conifold. In the case of ${\rm Im}\,z = +\infty$ the D4-brane is wrapped on $\mathcal{O}(-1)\to\mathbb{P}^1$, while in the case ${\rm Im}\,z =-\infty$ it is wrapped on the whole fiber directions $\mathbb{C}^2$. 

The BPS partition functions in these two large radius limits are evaluated in the field theories on the D4-brane wrapped on $\mathcal{O}(-1)\to \mathbb{P}^1$ and $\mathbb{C}^2$, respectively. In fact they were evaluated in \cite{AOSV} (see also \cite{Vafa-Witten, Vafa1}), which can be written in our notation as
\begin{eqnarray}
 Z_{+\infty}(u,v) &=& f(u)(1-v)\prod_{n=1}^\infty(1-u^n)(1-u^nv)(1-u^nv^{-1}), \label{eq:large1}
\\
Z_{-\infty}(u,v) &=& f(u)\prod_{n=1}^\infty(1-u^n), \label{eq:large2}
\end{eqnarray}
where $v$ and $u$ denote the chemical potentials for D2 and D0-branes, respectively. The function $f(u)$ is related to the bound states of D0-branes on the D4-brane without flux, which cannot be fixed because our D4-brane is non-compact. The degeneracy of the BPS bound states with non-vanishing D2-brane charge is unambiguously determined. When we consider the theory on a compact D4-brane, we find $f(u) = \prod_{n=1}^\infty (1-u^n)^{-\chi(C_4)}$ with the Euler characteristic of the 4-cycle $\chi(C_4)$.

The most interesting observation is that these two partition functions are related to each other via wall-crossing.
If we move the K\"ahler moduli from ${\rm Im}\,z = \infty$ to ${\rm Im}\,z = -\infty$, then various walls of marginal stability are crossed. When ${\rm Im}\,z =0$, the flop transition occurs and topology of the 4-cycle wrapped by the D4-brane is changed.\footnote{In order to keep the D2-branes massive, we tune the B-field for the rigid $\mathbb{P}^1$ so that ${\rm Re}\,z \not\in \mathbb{Z}$.} By using the Kontsevich-Soibelman wall-crossing formula, which we will briefly review in the next section, we find the relation between the partition functions in the two large radius limits, namely,
\begin{eqnarray}
 Z_{-\infty}(u,v) &=& Z_{+\infty}(u,v)\times \prod_{n=0}^\infty(1-u^nv)^{-1}\times \prod_{n=1}^\infty(1-u^nv^{-1})^{-1},
\end{eqnarray}
which perfectly matches Eqs. \eqref{eq:large1} and \eqref{eq:large2}. This means that the field theories on D4-branes in the two large radius limits can be connected by the wall-crossings involving the flop transition.

\section{Two D4-branes case}

We here study the wall-crossing of D4-D2-D0 bound states with two units of the same D4-brane charge. We introduce two non-compact D4-branes on the same supersymmetric divisor and consider D2-D0 bound states on them. In particular, we move the K\"ahler moduli $z$ for the rigid $\mathbb{P}^1$ and evaluate the jumps in the BPS partition function at the walls of marginal stability.

As in the case of one D4-brane, the moduli space is divided into two regions which are connected by the flop transition of the conifold. We put two D4-branes such that they are wrapped on $\mathcal{O}(-1)\to\mathbb{P}^1$ in the case of ${\rm Im}\,z > 0$. If we move to the region of ${\rm Im}\,z < 0$, the D4-branes are now wrapped on the whole fiber directions $\mathbb{C}^2$ and localized on the rigid $\mathbb{P}^1$ (see Fig. \ref{fig:flop}). The electric and magnetic charges of the BPS states of interest can be written as
\begin{eqnarray}
 \Gamma &=& 2\mathcal{D} + k\beta -ldV.
\end{eqnarray}

\subsection{Walls of marginal stability}

We here identify the walls of marginal stability in the moduli space. Since we now have two D4-branes, there are so-called ``primitive wall-crossings'' in addition to the ``semi-primitive wall-crossings.'' The former are associated with decays into two fragments of D4-D2-D0 bound states, while the latter are related to the separation of D2-D0 fragments. Note that in the case of one D4-brane we only have the semi-primitive wall-crossings.

For the BPS states with charge $\Gamma = 2\mathcal{D} + k\beta -ldV$, we consider decay channels of the form
$
\Gamma \to \Gamma_1  + \Gamma_2
$
where 
\begin{eqnarray}
\Gamma_1 &=& -a +(2-b)\mathcal{D} + (k-m)\beta -(l-n)dV,
\\
\Gamma_2 &=& a + b\mathcal{D} + m\beta - ndV.
\end{eqnarray}
Here $a,b,m$ and $n$ denote the D6, D4, D2 and D0 charges of $\Gamma_2$, respectively. The central charges of BPS states with charge $\Gamma_1$ and $\Gamma_2$ are evaluated as
\begin{eqnarray}
 Z(\Gamma_1) &=& -\frac{ac_6}{6}\Lambda^3e^{3i\varphi} - \frac{(2-b)c_4}{2}\Lambda^2e^{2i\varphi} + (k-m)z + (l-n),
\\
 Z(\Gamma_2) &=& \frac{ac_6}{6}\Lambda^3e^{3i\varphi} - \frac{bc_4}{2}\Lambda^2e^{2i\varphi} + mz + n.
\end{eqnarray}

We associate walls of marginal stability with these decay channels. They are defined as a subspace where $\arg[Z(\Gamma_1)] = \arg[Z(\Gamma_2)]$ is satisfied. It will turn out that there are only two types of walls in the finite $z$ region. First, if we assume $a\neq 0$, then $Z(\Gamma_1)$ and $Z(\Gamma_2)$ are dominated by the D6 and $\overline{\rm D6}$-brane contributions and never aligned. In fact, they are always {\em anti}-aligned and the corresponding BPS decay is forbidden in the local limit due to the energy conservation. So we assume $a=0$. Next, let us consider the case of $b\neq 0, 2$. In that case, $Z(\Gamma_1)$ and $Z(\Gamma_2)$ are dominated by the D4 (or $\overline{\rm D4}$) contribution and again anti-aligned unless $b=1$. We therefore assume $b=0$ or $b=1$. Note that the $b=2$ case is also included in the case of $b=0$ because we now have no D6-brane. Furthermore, in the case of $b=0$, it is sufficient to consider only the decay channels of $(m,n) = (\pm 1 ,n)$ since the only non-vanishing D2-D0 bound states on the conifold are known to be \eqref{eq:D2-D0}.

So from the above arguments, it turns out that there are the following two types of possible BPS decay channels:
\begin{eqnarray}
 \Gamma &\to& (\Gamma_1=\Gamma-\Gamma_2)  \;+\;  (\Gamma_2 = \mathcal{D} + m_2\beta-n_2dV), \label{eq:decay1}
\\
  \Gamma &\to&   (\Gamma_1 = \Gamma-\Gamma_2) \;+\; (\Gamma_2 = \pm\beta - NdV).
\label{eq:decay2}
\end{eqnarray}
The upper type is primitive decay and the lower one is semi-primitive decay. The walls of marginal stability are associated with these decay channels. We can identify the locations of these walls by solving the equation $\arg[Z(\Gamma_1)] = \arg[Z(\Gamma_2)]$.

For the decay channel \eqref{eq:decay1}, the location of the wall is given by
\begin{eqnarray}
 \frac{y}{\tan 2\varphi} &=& x + \frac{l-2n}{k-2m} \;\;=\;\; x + \frac{n_1-n_2}{m_1-m_2},
\end{eqnarray}
where $x$ and $y$ denote the real and imaginary part of the complexified K\"ahler parameter for the compact two cycle, namely, $z=x+iy$. Here we also define $m_1$ and $n_1$ so that $k=m_1+m_2$ and $l=n_1+n_2$. We also recall that $\varphi$ is related to the ratio of real and imaginary part of the K\"ahler parameter for non-compact cycles. We denote these walls as $V^{m_1,m_2}_{n_1,n_2}$.

\begin{figure}
\begin{center}
\includegraphics[width=12cm]{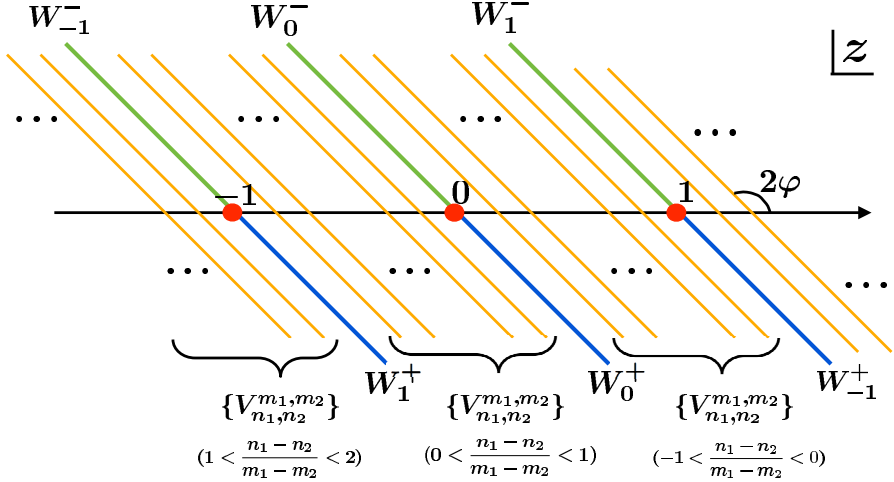}
\caption{Walls of marginal stability in $z$-plane with fixed $\varphi$. In the local limit, all the walls are straight lines whose slope is $2\varphi$. On the green and blue lines the moduli cross the walls $W_{N}^\pm$ as well as $V^{m_1,m_2}_{n_1,n_2}$ for $(n_1-n_2)/(m_1-m_2)\in\mathbb{Z}$. On the orange lines the other primitive walls $V^{m_1,m_2}_{n_1,n_2}$ are crossed. The red dots denote the singularities where D2-branes wrapped on $\mathbb{P}^1$ become massless. In this paper, we move the moduli without passing through these singularities.}\label{fig:walls2}
\end{center}
\end{figure}

Next, for the decay \eqref{eq:decay2} the position of the wall is given by
\begin{eqnarray}
 \frac{y}{\tan 2\varphi} &=& x \pm N.
\end{eqnarray}
We denote these walls as $W_N^\pm$. Note that the walls $\{V_{n_1,n_2}^{m_1,m_2}\}$ for  $(n_1-n_2)/(m_1-m_2) = \pm N$ are on the same position in the moduli space as the wall $W_N^\pm$ is. Thus, the walls of marginal stability can be depicted as in Fig. \ref{fig:walls2}

In the next subsection, we will evaluate the discrete change in BPS index that occurs when the moduli cross these walls.

\subsection{Wall-crossing formula for partition function}

We here study how the index of BPS states with charge $\Gamma = 2\mathcal{D} + k\beta -ldV$ is changed when the moduli cross the walls of marginal stability, by using the Kontsevich-Soibelman wall-crossing formula. When the Calabi-Yau moduli $t$ cross the walls of marginal stability, the degeneracy $\Omega(\gamma;t)$ changes. But the Kontsevich-Soibelman formula says that the product
\begin{eqnarray}
 A &=& \prod^{\longrightarrow}_{\gamma} U_\gamma^{\Omega(\gamma;t)},\label{KS-product}
\end{eqnarray}
which is taken in the decreasing order of $\arg Z(\gamma)$, is unchanged. Here $U_\gamma = \exp\sum_{n=1}^\infty \frac{1}{n^2} e_{n\gamma}$ is defined in terms of generators $e_{\gamma}$ of an infinite-dimensional Lie algebra with the commutation relation
\begin{eqnarray}
\left[e_{\gamma_1},\, e_{\gamma_2}\right] &=& \left(-1\right)^{\left<\gamma_1,\gamma_2\right>}\left<\gamma_1, \gamma_2\right> e_{\gamma_1+\gamma_2}.\label{eq:commutation}
\end{eqnarray}
The ordering of the product in eq. \eqref{KS-product} depends on the moduli $t$ as well as the index $\Omega(\gamma;t)$. The Kontsevich-Soibelman formula says that these two conpensate to keep $A$ invariant under the wall-crossing. We can read off the moduli dependence of $\Omega(\gamma;t)$ from this invariance of $A$ at the walls of marginal stability.

We will use this formula below to study the wall-crossings with respect to $V_{n_1,n_2}^{m_1,m_2}$ and $W_{N}^\pm$.

\subsubsection{Walls of $V_{n_1,n_2}^{m_1,m_2}$}

We first consider the walls $V_{n_1,n_2}^{m_1,m_2}$ for $(m_1,n_1;m_2,n_2)$ that has no integer $N$ satisfying $(n_1-n_2)/(m_1-m_2) = \pm N$. For such walls, we can use the primitive wall-crossing formula \cite{Denef-Moore}
\begin{eqnarray}
 \widetilde{\Omega}(\Gamma) &=& \Omega(\Gamma) + (-1)^{\left<\Gamma_1,\Gamma_2\right>}\left<\Gamma_1,\Gamma_2\right>\Omega(\Gamma_1)\Omega(\Gamma_2),
\end{eqnarray}
where $\Omega(\Gamma)$ and $\widetilde{\Omega}(\Gamma)$ denote the BPS index before the wall-crossing and after the wall-crossing, respectively. The physical meaning of this formula can be understood if we note that two-centered BPS solutions in four dimensions with charge $\Gamma_1$ and $\Gamma_2$ carry an intrinsic angular momentum $J = \frac{1}{2}(\left|\left<\Gamma_1,\Gamma_2\right>\right| -1)$ \cite{Denef:2002ru}. The sign factor $(-1)^{\left<\Gamma_1,\Gamma_2\right>} = -(-1)^{2J}$ is necessary because the BPS index has $(-1)^{2J}$ factor in the trace over the Hilbert space. This primitive wall-crossing formula can, of course, be derived from the Kontsevich-Soibelman formula.

\subsubsection{Walls of $W_N^\pm$}

We now study the wall-crossing with respect to $W_N^\pm$. We first notice that when the moduli cross $W_N^\pm$, the walls $\{V_{n_1,n_2}^{m_1,m_2}\}$ for $(n_1-n_2)/(m_1-m_2)=\pm N$ are also crossed simultaneously. So we use the full Kontsevich-Soibelman wall-crossing formula in order to read off the jump in the BPS index. We will use the following short hand notation below:
\begin{eqnarray}
\Gamma^{(b)}_{m,n} := b\mathcal{D} + m\beta -ndV,\quad \gamma_{\pm,N} := \pm \beta -NdV,
\end{eqnarray}
where $b=1,2$ and $m,n, N\in \mathbb{Z}$.

Suppose ${\rm Im}[Z(\Gamma^{(b)}_{m,n})\overline{Z(\gamma_{\pm,N})}]$ is positive before the wall-crossing and becomes negative after the moduli cross the wall. From the Kontsevich-Soibelman formula, we obtain the following equality:
\begin{eqnarray}
 \left(\prod_{\stackrel{b=1,2}{m,n\in \mathbb{Z}}}^{\longrightarrow}U_{\Gamma^{(b)}_{m,n}}^{\widetilde{\Omega}(\Gamma^{(b)}_{m,n})}\right)U_{\gamma_{\pm,N}}^{\widetilde{\Omega}(\gamma_{\pm,N})} &=& U_{j\gamma_{\pm,N}}^{\Omega(\gamma_{\pm,N})}\left(\prod_{\stackrel{b=1,2}{m,n\in\mathbb{Z}}}^{\longrightarrow}U_{\Gamma^{(b)}_{m,n}}^{\Omega(\Gamma^{(b)}_{m,n})}\right).\label{eq:semi-primitive}
\end{eqnarray}
Since we know $\Omega(\gamma_{\pm,N})$ has no wall-crossing on the conifold \cite{Nishinaka:2010qk}, it follows that
\begin{eqnarray}
 \widetilde{\Omega}(\gamma_{\pm,N}) &=& \Omega(\gamma_{\pm,N}).
\end{eqnarray}
Then by multiplying \eqref{eq:semi-primitive} by $U^{-\Omega(\gamma_{\pm,N})}_{\gamma_{\pm,N}}$ from the right, we obtain
\begin{eqnarray}
  \left(\prod_{\stackrel{b=1,2}{m,n\in \mathbb{Z}}}^{\longrightarrow}U_{\Gamma^{(b)}_{m,n}}^{\widetilde{\Omega}(\Gamma^{(b)}_{m,n})}\right) &=& U_{j\gamma_{\pm,N}}^{\Omega(\gamma_{\pm,N})}\left(\prod_{\stackrel{b=1,2}{m,n\in\mathbb{Z}}}^{\longrightarrow}U_{\Gamma^{(b)}_{m,n}}^{\Omega(\Gamma^{(b)}_{m,n})}\right)U_{\gamma_{\pm,N}}^{-\Omega(\gamma_{\pm,N})}.\label{eq:UUU}
\end{eqnarray}
From the above equality, we can in principle read off the discrete change in the BPS index $\Omega(\Gamma^{(2)}_{m,n})$ for the BPS states with two units of D4-brane charge because we already know the full moduli dependence of $\Omega(\Gamma^{(1)}_{m,n})$, which is the index for BPS states with {\rm one} unit of D4 charge, from the result in \cite{Nishinaka:2010qk}. In fact, by a short calculation shown in Appendix A, we can derive the following equation from Eq. \eqref{eq:UUU}:
\begin{eqnarray}
&& \sum_{m,n}\left\{\widetilde{\Omega}(\Gamma^{(2)}_{m,n})e_{\Gamma^{(2)}_{m,n}} + \frac{1}{4}\widetilde{\Omega}(\Gamma^{(1)}_{m,n})e_{2\Gamma^{(1)}_{m,n}}\right\}
\nonumber \\
&& + \;1/2 \!\!\!\!\!\!\!\!\!\!\!\!\!\!\!\!\!\!\!\!\!\!\!\!\sum_{\arg\widetilde{Z}(\Gamma^{(1)}_{m_1,n_1})>\arg\widetilde{Z}(\Gamma^{(1)}_{m_2,n_2})}\!\!\!\!\!\!\!\!\!\!\!\!\!\!\!\!\!\!\!\!\!
(-1)^{\left<\Gamma^{(1)}_{m_1,n_1},\Gamma^{(1)}_{m_2,n_2}\right>}\left<\Gamma^{(1)}_{m_1,n_1},\Gamma^{(1)}_{m_2,n_2}\right>\widetilde{\Omega}(\Gamma^{(1)}_{m_1,n_1})\widetilde{\Omega}(\Gamma^{(1)}_{m_2,n_2})e_{\Gamma^{(1)}_{m_1,n_1}+\Gamma^{(1)}_{m_2,n_2}}
\nonumber \\[3mm]
 &=&
\sum_{m,n}\left\{\Omega(\Gamma^{(2)}_{m,n})\hat{e}_{\Gamma^{(2)}_{m,n}} + \frac{1}{4}\Omega(\Gamma^{(1)}_{m,n})\hat{e}_{2\Gamma^{(1)}_{m,n}} \right\}
\nonumber \\
&&
+ \; 1/2 \!\!\!\!\!\!\!\!\!\!\!\!\!\!\!\!\!\!\!\!\!\!\!\! \sum_{{\arg Z(\Gamma^{(1)}_{m_1,n_1}) > \arg Z(\Gamma^{(1)}_{m_2,n_2})}} \!\!\!\!\!\!\!\!\!\!\!\!\!\!\!\!\!\!\!\!\!(-1)^{\left<\Gamma^{(1)}_{m_1,n_1},\Gamma^{(1)}_{m_2,n_2}\right>} \left<\Gamma^{(1)}_{m_1,n_1},\Gamma^{(1)}_{m_2,n_2}\right>\Omega(\Gamma^{(1)}_{m_1,n_1})\Omega(\Gamma^{(1)}_{m_2,n_2})\hat{e}_{\Gamma^{(1)}_{m_1,n_1}+\Gamma^{(1)}_{m_2,n_2}},
\nonumber \\
\label{eq:WCF2}
\end{eqnarray}
where we used a short hand notation
\begin{eqnarray}
 \hat{e}_{\Gamma} &=& U^{\Omega(\gamma_{\pm,N})}_{\gamma_{\pm,N}}\,e_{\Gamma}\;U^{-\Omega(\gamma_{\pm,N})}_{\gamma_{\pm,N}}.\label{eq:ueu}
\end{eqnarray}
We also denote by $Z(\Gamma)$ and $\widetilde{Z}(\Gamma)$ the central charges before the wall-crossing and after the wall-crossing, respectively.
Here the last sum in the left-hand side of \eqref{eq:WCF2} runs over all possible $(m_1,n_1;m_2,n_2)\in\mathbb{Z}^4$ satisfying the inequality $\arg[\widetilde{Z}(\Gamma^{(1)}_{m_1,n_1})] > \arg[\widetilde{Z}(\Gamma^{(1)}_{m_2,n_2})]$. The last sum in the right-hand side is similar. We also note for the later use that \eqref{eq:ueu} can be rewritten as
\begin{eqnarray}
 \hat{e}_{\Gamma} &=& e_{\Gamma}\circ \left\{1 + (-1)^{\left<\Gamma,\gamma_{\pm,N}\right>}e_{\gamma_{\pm,N}}\right\}^{\circ \left\{-\left<\Gamma,\;\gamma_{\pm,N}\right>\Omega(\gamma_{\pm,N})\right\}},
\end{eqnarray}
where $\circ$ denotes a commutative product $e_{\gamma_1}\circ e_{\gamma_2} := e_{\gamma_1+\gamma_2}$.

The first term in the bracket in Eq. \eqref{eq:WCF2} is clearly related to the partition function of BPS states with two units of D4 charge, while the last sum is associated with the primitive bound states of D4-D2-D0 fragments. The second term in the bracket is somewhat strange but by defining the rational invariants by 
\begin{eqnarray}
 \overline{\Omega}(\Gamma) &=& \sum_{m|\Gamma}\frac{\Omega(\Gamma/m)}{m^2},
\end{eqnarray}
we find that Eq. \eqref{eq:WCF2} shows the jumps in the rational invariants $\Omega(\Gamma^{(2)}_{m,n})$ can be understood by the contributions from the split flow trees \cite{Manschot2}.

From Eq.~\eqref{eq:WCF2} we can in principle read off the discrete change in the BPS index $\Omega(\Gamma^{(2)}_{m,n})$ for the BPS states with two units of D4-brane charge. This is, however, a rather hard task and we will not perform such a calculation for each wall-crossing. But if we compare two large radius limits in the moduli space, then the above calculations lead to a suggestive result, which will be discussed in the next subsection.

\subsection{Two large radius limits}

In the previous subsection, we have explained how we can read off the jumps in the BPS index of interest $\Omega(\Gamma^{(2)}_{m,n})$ from the Kontsevich-Soibelman wall-crossing formula. We now compare the BPS index in the two large radius limits of the rigid $\mathbb{P}^1$ by using the similar arguments as in the previous subsection.

As was mentioned before, we have two large radius limits in the moduli space, namely ${\rm Im}\,z = \pm \infty$. In these limits, the BPS index is expected to be evaluated in the field theory on D4-branes. We can compare the BPS indices in these two limits by using the Kontsevich-Soibelman wall-crossing formula. Suppose that we fix the real part of the moduli so that ${\rm Re}\,z = 1/2$ and move the imaginary part from ${\rm Im}\,z =+\infty$ to ${\rm Im}\,z = -\infty$.   We denote by $\Omega_{\pm\infty}(\Gamma^{(2)}_{m,n})$ the index of BPS states with charge $\Gamma^{(2)}_{m,n}$ in the limits ${\rm Im}\,z = \pm\infty$, respectively. Then by the similar argument as in the previous subsection, the Kontsevich-Soibelman wall-crossing formula leads to the equality of
\begin{eqnarray}
&& \sum_{m,n}\overline{\Omega}_{-\infty}(\Gamma^{(2)}_{m,n})e_{\Gamma^{(2)}_{m,n}}
 \;\; \; + \;\;\;1/2\!\!\!\!\!\!\!\!\!\!\!\!\!\!\!\!\!\!\!\!\!\!\!\!\!\!\!\sum_{\arg Z_{-\infty}(\Gamma^{(1)}_{m_1,n_1})>\arg Z_{-\infty}(\Gamma^{(1)}_{m_2,n_2})}\!\!\!\!\!\!\!\!\!\!\!\!\!\!\!\!\!\!\!\!\!\!\!\!\!\!\!\!\left\{
(-1)^{\left<\Gamma^{(1)}_{m_1,n_1},\Gamma^{(1)}_{m_2,n_2}\right>}\left<\Gamma^{(1)}_{m_1,n_1},\Gamma^{(1)}_{m_2,n_2}\right>\right.
\nonumber \\[3mm]
&&\left.  \qquad\qquad \times \; \overline{\Omega}_{-\infty}(\Gamma^{(1)}_{m_1,n_1})\overline{\Omega}_{-\infty}(\Gamma^{(1)}_{m_2,n_2})e_{\Gamma^{(1)}_{m_1,n_1}+\Gamma^{(1)}_{m_2,n_2}}\right\}\label{eq:pm-infty1-1}
\end{eqnarray}
and
\begin{eqnarray}
&&\sum_{m,n}\overline{\Omega}_{+\infty}(\Gamma^{(2)}_{m,n})\hat{e}_{\Gamma^{(2)}_{m,n}}
\;\;\;+ \;\;\; 1/2 \!\!\!\!\!\!\!\!\!\!\!\!\!\!\!\!\!\!\!\!\!\!\!\!\!\!\! \sum_{{\arg Z_{+\infty}(\Gamma^{(1)}_{m_1,n_1}) > \arg Z_{+\infty}(\Gamma^{(1)}_{m_2,n_2})}} \!\!\!\!\!\!\!\!\!\!\!\!\!\!\!\!\!\!\!\!\!\!\!\!\!\!\!\!\left\{(-1)^{\left<\Gamma^{(1)}_{m_1,n_1},\Gamma^{(1)}_{m_2,n_2}\right>} \left<\Gamma^{(1)}_{m_1,n_1},\Gamma^{(1)}_{m_2,n_2}\right>\right.
\nonumber \\[3mm]
&&\left. \qquad\qquad \times \;\overline{\Omega}_{+\infty}(\Gamma^{(1)}_{m_1,n_1})\overline{\Omega}_{+\infty}(\Gamma^{(1)}_{m_2,n_2})\hat{e}_{\Gamma^{(1)}_{m_1,n_1} + \Gamma^{(1)}_{m_2,n_2}}\right\}. \label{eq:pm-infty1-2}
\end{eqnarray}
Here it follows that
\begin{eqnarray}
 \left<\Gamma^{(1)}_{m_1,n_1},\Gamma^{(1)}_{m_2,n_2}\right> &=& m_1-m_2,
\\[2mm]
\hat{e}_{\Gamma^{(2)}_{m,n}} &=& e_{\Gamma^{(2)}_{m,n}}\circ\left\{\prod_{n=0}^\infty \circ (1+e_{\gamma_{+,n}})^{\circ (-2)}\right\} \circ \left\{\prod_{n=1}^\infty \circ(1+e_{\gamma_{-,n}})^{\circ (-2)}\right\},
\end{eqnarray}
with the previously defined commutative product $e_{\gamma_1}\circ e_{\gamma_2} = e_{\gamma_1+\gamma_2}$.
By replacing $e_{\Gamma^{(2)}_{m,n}}$ with Boltzmann factors $v^{m}u^n$, the equality of \eqref{eq:pm-infty1-1} and \eqref{eq:pm-infty1-2} implies that
\begin{eqnarray}
&&\!\!\!\!\!\!\!\! \mathcal{Z}^{(2)}_{-\infty}(u,v)
\nonumber \\[2mm]
&&\!\!\!\! +  \;\;\;1/2\!\!\!\!\!\!\!\!\!\!\!\!\!\!\!\!\!\!\!\!\!\!\!\!\!\!\!\!\! \sum_{{\arg Z_{-\infty}(\Gamma^{(1)}_{m_1,n_1}) > \arg Z_{-\infty}(\Gamma^{(1)}_{m_2,n_2})}} \!\!\!\!\!\!\!\!\!\!\!\!\!\!\!\!\!\!\!\!\!\!\!\!\!\!\!\!(-1)^{m_1-m_2}(m_1-m_2)\overline{\Omega}_{-\infty}(\Gamma^{(1)}_{m_1,n_1})\overline{\Omega}_{-\infty}(\Gamma^{(1)}_{m_2,n_2})v^{m_1+m_2}u^{n_1+n_2}
\nonumber \\[5mm]
&=& \prod_{n=0}^\infty(1+u^rv)^{-2}\prod_{n=1}^\infty(1+u^rv^{-1})^{-2}\times \Biggl\{\mathcal{Z}^{(2)}_{+\infty}(u,v)\Biggr.
\nonumber \\[-1mm]
&&\Biggl.\;\; +  \;\;\;1/2\!\!\!\!\!\!\!\!\!\!\!\!\!\!\!\!\!\!\!\!\!\!\!\!\!\!\!\!\! \sum_{{\arg Z_{+\infty}(\Gamma^{(1)}_{m_1,n_1}) > \arg Z_{+\infty}(\Gamma^{(1)}_{m_2,n_2})}} \!\!\!\!\!\!\!\!\!\!\!\!\!\!\!\!\!\!\!\!\!\!\!\!\!\!\!\! (-1)^{m_1-m_2}(m_1-m_2)\overline{\Omega}_{+\infty}(\Gamma^{(1)}_{m_1,n_1})\overline{\Omega}_{+\infty}(\Gamma^{(1)}_{m_2,n_2})v^{m_1+m_2}u^{n_1+n_2}\Biggr\},
\nonumber \\
\label{eq:pm-infty2}
\end{eqnarray}
with generating functions of the rational invariants defined by
\begin{eqnarray}
 \mathcal{Z}^{(b)}_{\pm\infty}(u,v) &=& \sum_{m,n}\overline{\Omega}_{\pm\infty}(\Gamma^{(b)}_{m,n})\,v^mu^n.
\end{eqnarray}

We now recall the explicit expression of the generating functions for the BPS states with {\em one} unit of D4-brane charge\footnote{Note that for BPS states with one unit of D4-brane charge $\overline{\Omega}(\Gamma^{(1)}_{m,n}) = \Omega(\Gamma^{(1)}_{m,n})$ follows.}
\begin{eqnarray}
\mathcal{Z}^{(1)}_{+\infty}(u,v) &=& f(u)(1-v)\prod_{n=1}^\infty(1-u^n)(1-u^nv)(1-u^nv^{-1}),\label{eq:1p-infty}
\\
\mathcal{Z}^{(1)}_{-\infty}(u,v) &=& f(u)\prod_{n=1}^\infty(1-u^r).
\end{eqnarray}
In particular, we should note that $Z_{-\infty}^{(1)}(u,v)$ has no $v$-dependence. Thus, $\overline{\Omega}_{-\infty}(\Gamma^{(1)}_{m,n})$ with non-zero $m$ turns to be zero and the second term in the left-hand side of Eq. \eqref{eq:pm-infty2} vanishes. The second term in the right-hand side can also be read off by expanding Eq. \eqref{eq:1p-infty} as
\begin{eqnarray}
\mathcal{Z}^{(1)}_{+\infty}(u,v) &=& f(u)\sum_{n\in\mathbb{Z}}(-1)^{n}u^{\frac{n(n-1)}{2}}v^n.
\end{eqnarray}
Furthermore, the condition $\arg[Z(\Gamma^{(1)}_{m_1,n_1})]>\arg[Z(\Gamma^{(1)}_{m_2,n_2})]$ now turns to be $m_1>m_2$ because of the limit of ${\rm Im}\,z = +\infty$.
Then we finally obtain
\begin{eqnarray}
\mathcal{Z}^{(2)}_{-\infty}(u,v) &=& \frac{\mathcal{Z}^{(2)}_{+\infty}(u,v) + \frac{1}{2}[f(u)]^2\sum_{m_1>m_2}(m_1-m_2)u^{\frac{m_1(m_1-1)}{2}+\frac{m_2(m_2-1)}{2}}v^{m_1+m_2}}{\prod_{n=0}^\infty(1+u^nv)^{2}\prod_{n=1}^\infty(1+u^nv^{-1})^{2}}.
\nonumber \\ \label{eq:relation}
\end{eqnarray}
This is a suggestive result. The generating function of rational invariants with two units of D4-brane charge suffers from discrete changes with respect to primitive wall-crossings and semi-primitive wall-crossings when the moduli move from ${\rm Im}\,z = +\infty$ to ${\rm Im}\,z = -\infty$. The second term in the numerator is associated with the primitive wall-crossings and the denominator describes the semi-primitive wall-crossings.

\subsubsection{Two-centered states at large radius}

We now study the explicit expression of $\mathcal{Z}_{+\infty}^{(2)}(u,v)$. In the limit ${\rm Im}\,z = +\infty$, the D4-branes are wrapped on $\mathcal{O}(-1)\to\mathbb{P}^1$. Therefore, one might expect that $\mathcal{Z}_{+\infty}^{(2)}(u,v)$ is equivalent to the partition function for the field theory on two D4-branes wrapped on $\mathcal{O}(-1)\to\mathbb{P}^1$. Such a partition function was evaluated in \cite{AOSV}. In our notation, it can be written as
\begin{eqnarray}
\mathcal{Z}_{\rm field}^{(2)}(u,v) &=&  [f(u)]^2(1+v)^2\prod_{n=1}^\infty (1-u^n)^2(1+u^nv)^2(1+u^nv^{-1})^2.
\end{eqnarray}
This looks like the square of the partition function for one D4-brane \eqref{eq:1p-infty}, but we here added an additional minus sign to the D2-brane chemical potential $v$. This sign comes from the fact that the bound state of a D2-brane and $N$ D4-branes wrapped on $\mathcal{O}(-1)\to\mathbb{P}^1$ has an intrinsic angular momentum $1/2(\left<\beta,N\mathcal{D}\right> -1)= -1/2(N+1)$.

If this is the correct expression for $\mathcal{Z}_{+\infty}^{(2)}(u,v)$, then by using the relation \eqref{eq:relation} we find that $\mathcal{Z}_{-\infty}^{(2)}(u,v)$ has a rather strange expression. In particular, $\mathcal{Z}_{-\infty}^{(2)}(u,v)$ has a non-trivial $v$-dependence even though it is a generating function of the BPS (rational) indices in the large radius limit. In the large radius limit, we expect that the BPS index can be counted in the field theory on D4-branes, now wrapped on the whole fiber direction $\mathbb{C}^2$ and localized on the rigid $\mathbb{P}^1$ in the conifold. Since the D4-branes are localized on $\mathbb{P}^1$, the flux on the D4-branes cannot induce any D2-brane charge. Thus, we should expect that $\mathcal{Z}_{-\infty}^{(2)}(u,v)$ has no D2-brane charge contribution and is independent of $v$. This consideration leads us to the following observation
\begin{eqnarray}
 \mathcal{Z}_{+\infty}^{(2)}(u,v) &=& [f(u)]^2(1+v)^2\prod_{n=1}^\infty (1-u^n)^2(1+u^nv)^2(1+u^nv^{-1})^2
\nonumber \\
&& \quad -\; \frac{1}{2}[f(u)]^2\sum_{m_1>m_2}(m_1-m_2)u^{\frac{m_1(m_1-1)}{2}+\frac{m_2(m_2-1)}{2}}v^{m_1+m_2}, \label{eq:p-infty}
\\[2mm]
\mathcal{Z}_{-\infty}^{(2)}(u,v) &=& [f(u)]^2\prod_{n=1}^\infty(1-u^n)^2. \label{eq:m-infty}
\end{eqnarray}
Note that this expression for $\mathcal{Z}^{(2)}_{-\infty}$ is independent of $v$. If we recall that $f(u) = \prod_{n=1}^\infty(1-u^n)^{-\chi(C_4)}$ for a compact Calabi-Yau case, then this is also consistent with the fact that the Euler characteristic decreases by one through the flop transition of the conifold.

In \eqref{eq:p-infty}, the first term in the right-hand side is the previous partition function of the field theory on the D4-branes wrapped on $\mathcal{O}(-1)\to\mathbb{P}^1$. On the other hand, the second term means that there are two-centered bound states of D4-D2-D0 fragments even in the limit of ${\rm Im}\,z = +\infty$.
We can show this by supergravity analysis. In four-dimensional supergravity, a two-centered BPS solution with charge $\Gamma_1$ and $\Gamma_2$ for each center has the following distance between centers:
\begin{eqnarray}
 R &=& \frac{\left<\Gamma_1,\Gamma_2\right>}{2}\frac{\left|Z(\Gamma_1)+Z(\Gamma_2)\right|}{{\rm Im}(Z(\Gamma_1)\overline{Z(\Gamma_2)})}.
\end{eqnarray}
Since this distance should be positive, the inequality
\begin{eqnarray}
 \left<\Gamma_1,\Gamma_2\right>{\rm Im}(Z(\Gamma_1)\overline{Z(\Gamma_2)}) &>& 0 \label{eq:condition}
\end{eqnarray}
follows in the stable side of the walls of marginal stability \cite{Denef-Moore}. So we now check that this inequality is satisfied in the limit of ${\rm Im}\,z = +\infty$. The electromagnetic charges of two centers of interest are $\Gamma_1 = \mathcal{D}+m_1\beta-n_1dV$ and $\Gamma_2 = \mathcal{D}+m_2\beta-n_2dV$, respectively. The charge intersection product is
\begin{eqnarray}
 \left<\Gamma_1,\Gamma_2\right> &=& m_1-m_2.
\end{eqnarray}
On the other hand, the central charges are evalutated as
\begin{eqnarray}
 Z(\Gamma_1) = -\frac{c_4}{2}\Lambda^2e^{2i\varphi} + m_1z + n_1,\qquad
 Z(\Gamma_2) = -\frac{c_4}{2}\Lambda^2e^{2i\varphi} + m_2z + n_2,
\end{eqnarray}
and therefore
\begin{eqnarray}
{\rm Im}(Z(\Gamma_1)\overline{Z(\Gamma_2)}) &=& -c_4\Lambda^2{\rm Im}\left[e^{-2i\varphi}(m_1z + n_1) + e^{2i\varphi}(m_2\bar{z}+n_2)\right].
\end{eqnarray}
If we recall $c_4>0$ and $\pi/2 >\varphi >\pi/4$, we find that \eqref{eq:condition} holds for all $(m_1,n_1;m_2,n_2)\in \mathbb{Z}^4$ in the limit of ${\rm Im}\,z = +\infty$ and  all the primitive bound states of D4-D2-D0 fragments are stable in the limit of ${\rm Im}\,z=+\infty$.\footnote{In general, the bound states might be unstable even when the inequality \eqref{eq:condition} holds. The reason for this is that if the moduli cross the {\em walls of anti-aligned}, where $Z(\Gamma_1)$ and $Z(\Gamma_2)$ are anti-aligned and ${\rm Im}(Z(\Gamma_1)\overline{Z(\Gamma_2)})$ vanishes, then $\left<\Gamma_1,\Gamma_2\right>{\rm Im}(Z(\Gamma_1)\overline{Z(\Gamma_2)})$ changes its sign even though there is no jump in the BPS indices at the wall. But since we are now considering BPS states with at least one non-compact D4-brane charge, there is no wall of anti-aligned in the local limit of the conifold. Thus \eqref{eq:condition} immediately implies that the bound states are stable.} This means that  $\mathcal{Z}_{+\infty}^{(2)}(u,v)$ has contributions from two-centered D4-D2-D0 bound states, which is consistent with Eq. \eqref{eq:p-infty}. 

One might think that this leads to a contradiction with the fact that the entropy of {\em single-centered} black holes with D4-D2-D0 charges can be evaluated in MSW CFT in the large radius and large charge limit, which is the low-energy effective theory of wrapped M5-branes \cite{MSW}. In fact, our observations show that even in the large radius limit of ${\rm Im}\,z = +\infty$ the BPS wrapped D4-D2-D0 system has {\em multi-centered} constituents. However, if the moduli $t$ are fixed at the attractor point $t_{\rm att}(\Gamma)$ for a given electromagnetic charge $\Gamma$, then the index $\Omega(\Gamma;t_{\rm att}(\Gamma))$ has no multi-centered contributions, while the index for other charge $\Omega(\Gamma';t_{\rm att}(\Gamma))$ might have multi-centered contributions \cite{Manschot1,Manschot2}. We can show that the index $\Omega(\Gamma;t_{\rm att}(\Gamma))$ has no multi-centered contributions by using the attractor equation for BPS states with charge $\Gamma$
\begin{eqnarray}
 2{\rm Im}(\overline{Z(\Gamma;t)}\;{\bf \Omega}) &=& -\Gamma,\label{eq:attractor}
\end{eqnarray}
where ${\bf \Omega}$ denotes the normalized period vector. Solving this equation, we obtain the attractor point $t = t_{\rm att}(\Gamma)$ for the charge $\Gamma$. If we consider the BPS bound states with charge $\Gamma_1$ and $\Gamma_2$ for $\Gamma_1+\Gamma_2 = \Gamma$, the stable side of the walls of marginal stability is determined by the inequality \eqref{eq:condition}. But we can show that the attractor point $t_{\rm att}(\Gamma)$ is always in the unstable side of the wall for an arbitrary decay channel. In fact, by acting $\left<\Gamma_1\,,\;\cdot\;\;\right>$ on \eqref{eq:attractor} from the left and recalling $Z(\Gamma) = \left<\Gamma,{\bf \Omega}\right>$, we find
\begin{eqnarray}
 2{\rm Im}(\overline{Z(\Gamma_2)}Z(\Gamma_1)) &=& -\left<\Gamma_1,\Gamma_2\right>
\end{eqnarray}
at the attractor point $t=t_{\rm att}(\Gamma)$, which implies that the attractor point $t_{\rm att}(\Gamma)$ for the charge $\Gamma$ does not satisfy \eqref{eq:condition} for any $\Gamma_1$ and $\Gamma_2$.

\section{Discussions}

In this paper we studied the wall-crossing of D4-D2-D0 bound states on the conifold that have two units of D4-brane charge. We identified the walls of marginal stability and evaluated the discrete changes that occurs when the moduli crossing the walls. There are primitive wall-crossings for decays into two fragments of D4-D2-D0 and semi-primitive wall-crossings for the separation of D2-D0 fragments.

In particular, we evaluated the difference between the BPS partition functions in two large radius limits. The result \eqref{eq:relation} leads us to the observation that there are many two-centered bound states of two D4-D2-D0 fragments in the limit of ${\rm Im}\,z = +\infty$, which was shown by supergravity analysis. By taking into account these two-centered bound states, we observed that the field theories on D4-branes in the two large radius limits can be connected by wall-crossing involving the flop transition of the conifold, even in the case of two units of D4-brane charge.

The slight difference from the one D4-brane case is that we replaced the generating function of the integer indices $\Omega(\Gamma^{(2)}_{m,n})$ with that of the rational invariants $\overline{\Omega}(\Gamma^{(2)}_{m,n})$. This replacement has already been discussed in \cite{Manschot2} where it was deduced that the S-duality invariance of the generating function requires such a replacement. It seems interesting to study the physical meaning of this replacement further.

The generalization to the $N$ D4-branes case is also interesting for future work.  When we have $N$ D4-branes on $\mathcal{O}(-1)\to\mathbb{P}^1$, there are semi-primitive walls as well as many kinds of primitive walls. Then the chamber structure in the moduli space will be very complicated. Nevertheless, from the observation in this paper, we expect that by considering the generating function of rational invariants $\overline{\Omega}(\Gamma^{(N)}_{m,n})$ we can derive the similar relation to \eqref{eq:relation} that tells us the generating functions in two large radius limits are properly connected by wall-crossings.

\begin{figure}
\begin{center}
\includegraphics[width=3.5cm]{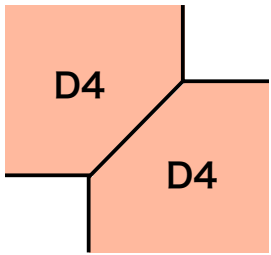}
\caption{Another setup with two units of D4-brane charge. The two D4-branes are now wrapped on different divisors and overlap over the rigid $\mathbb{P}^1$.}\label{fig:conifold3}
\end{center}
\end{figure}
We should also comment on the another setup with multiple D4-D2-D0 bound states. In this paper we put two non-compact D4-branes on the same supersymmetric divisor. But we can put them on different divisors. For example, we can put two D4-branes as in Fig. \ref{fig:conifold3}. It is interesting to study wall-crossing phenomena in this setup, which is related to the work of \cite{Jafferis-Saulina} where the wall-crossing of D4-D2-D0 fragments wrapped on ample divisors and overlapping over a compact Riemann surface are discussed in the large volume limit.

For other future directions, it is interesting to study the relation between the D4-D2-D0 wall-crossings and an infinite dimensional Lie algebra. Such a relation can be found in \cite{Cheng-Verlinde} for ${\mathcal N}=4$ string theory. The character of the affine Lie algebra appears in the instanton counting problem \cite{Nakajima, Vafa-Witten}. It is also interesting to make a statistical model like the crystal melting model for the D6-D4-D2-D0 system. The representation of the D4 partition function in terms of a sum over Young diagrams is well-known. Another example of such a statistical model is one in \cite{Jafferis:2006ny}.


\section*{Acknowledgments}

We would like to thank Takahiro Kubota and Satoshi Yamaguchi for many illuminating discussions, important comments and suggestions.
The author was supported in part by JSPS Research Fellowship for Young Scientists. The author thanks the Yukawa Institute for Theoretical Physics at Kyoto University, where this work was initiated during the YITP-W-10-02 on ``Development of Quantum Field Theory and String Theory.''

\begin{appendix}

\section{Wall-crossing with respect to $W_{N}^\pm$}

We here show the explicit derivation of Eq. \eqref{eq:WCF2} from Eq. \eqref{eq:UUU}. We first expand the exponentials in $U$s and collect all terms with two units of the D4-brane charge. Then the left-hand side of Eq. \eqref{eq:UUU} can be written as
\begin{eqnarray}
 &&\sum_{m,n}\widetilde{\Omega}(\Gamma^{(2)}_{m,n})e_{\Gamma^{(2)}_{m,n}} + \frac{1}{2!}\sum_{m,n}(\widetilde{\Omega}(\Gamma^{(1)}_{m,n})e_{\Gamma^{(1)}_{m,n}})^2 + \frac{1}{2^2}\sum_{m,n}\widetilde{\Omega}(\Gamma^{(1)}_{m,n})e_{2\Gamma^{(1)}_{m,n}} 
\nonumber \\[2mm]
&&+ \!\!\!\!\!\!\!\!\!\!\!\! \sum_{\arg \widetilde{Z}(\Gamma^{(1)}_{m_1,n_1}) > \arg \widetilde{Z}(\Gamma^{(1)}_{m_2,n_2})} \!\!\!\!\!\!\!\!\!\!\!\!\!\! \widetilde{\Omega}(\Gamma^{(1)}_{m_1,n_1})\widetilde{\Omega}(\Gamma^{(1)}_{m_2,n_2})e_{\Gamma^{(1)}_{m_1,n_1}}e_{\Gamma^{(1)}_{m_2,n_2}}\label{eq:left}
\end{eqnarray}
and, on the other hand, the right-hand side becomes
\begin{eqnarray}
&& U_{\gamma_{\pm,N}}^{\Omega(\gamma_{\pm,N})}
\left\{
\sum_{m,n}\Omega(\Gamma^{(2)}_{m,n})e_{\Gamma^{(2)}_{m,n}} + \frac{1}{2!}\sum_{m,n}(\Omega(\Gamma^{(1)}_{m,n})e_{\Gamma^{(1)}_{m,n}})^2 + \frac{1}{2^2}\sum_{m,n}\Omega(\Gamma^{(1)}_{m,n})e_{2\Gamma^{(1)}_{m,n}} 
\right.
\nonumber \\
&&\left. 
 \qquad\qquad
+ \!\!\!\!\!\!\!\!\!\!\!\! \sum_{{\arg Z(\Gamma^{(1)}_{m_1,n_1}) > \arg Z(\Gamma^{(1)}_{m_2,n_2})}} \!\!\!\!\!\!\!\!\!\!\!\!\!\! \Omega(\Gamma^{(1)}_{m_1,n_1})\Omega(\Gamma^{(1)}_{m_2,n_2})e_{\Gamma^{(1)}_{m_1,n_1}}e_{\Gamma^{(1)}_{m_2,n_2}}
\right\}
U_{\gamma_{\pm,N}}^{-\Omega(\gamma_{\pm,N})}.\label{eq:right}
\end{eqnarray}
Here the last sum in \eqref{eq:left} runs over all possible $(m_1,n_1;m_2.n_2)\in\mathbb{Z}^4$ satisfying the inequality $\arg[\widetilde{Z}(\Gamma^{(1)}_{m_1,n_1})] > \arg[\widetilde{Z}(\Gamma^{(1)}_{m_2,n_2})]$. The last sum in \eqref{eq:right} is similar. We should note that some charge combinations $(m_1,n_1;m_2,n_2)$ satisfy  $\arg[\widetilde{Z}(\Gamma^{(1)}_{m_1,n_1})] > \arg[\widetilde{Z}(\Gamma^{(1)}_{m_2,n_2})]$ but do not satisfy  $\arg[Z(\Gamma^{(1)}_{m_1,n_1})] > \arg[Z(\Gamma^{(1)}_{m_2,n_2})]$. Such combinations are associated with the primitive walls $V_{n_1,n_2}^{m_1,m_2}$ that are simultaneously crossed with the wall $W_{N}^\pm$. For such $(m_1,n_1;m_2,n_2)$, the order of $e_{\Gamma^{(1)}_{m_1,n_1}}$ and $e_{\Gamma^{(1)}_{m_2,n_2}}$ is reversed when the moduli cross the wall.

By using the identity
\begin{eqnarray}
 e_{\gamma_1}e_{\gamma_2} &=& \frac{1}{2}\left\{e_{\gamma_1},e_{\gamma_2}\right\} + \frac{1}{2}\left[e_{\gamma_1},e_{\gamma_2}\right],
\end{eqnarray}
and the commutation relation \eqref{eq:commutation}, we can rewrite \eqref{eq:left} as
\begin{eqnarray}
&& \sum_{m,n}\left\{\widetilde{\Omega}(\Gamma^{(2)}_{m,n})e_{\Gamma^{(2)}_{m,n}} + \frac{1}{4}\widetilde{\Omega}(\Gamma^{(1)}_{m,n})e_{2\Gamma^{(1)}_{m,n}}\right\} + \frac{1}{2}\left\{\sum_{m,n}\widetilde{\Omega}(\Gamma^{(1)}_{m,n})e_{\Gamma^{(1)}_{m,n}}\right\}^2
\nonumber \\
&& + \frac{1}{2}\!\!\!\!\!\!\!\!\!\!\!\!\!\!\!\!\!\!\!\sum_{\arg\widetilde{Z}(\Gamma^{(1)}_{m_1,n_1})>\arg\widetilde{Z}(\Gamma^{(1)}_{m_2,n_2})}\!\!\!\!\!\!\!\!\!\!\!\!\!\!\!\!\!\!
(-1)^{\left<\Gamma^{(1)}_{m_1,n_1},\Gamma^{(1)}_{m_2,n_2}\right>}\left<\Gamma^{(1)}_{m_1,n_1},\Gamma^{(1)}_{m_2,n_2}\right>\widetilde{\Omega}(\Gamma^{(1)}_{m_1,n_1})\widetilde{\Omega}(\Gamma^{(1)}_{m_2,n_2})e_{\Gamma^{(1)}_{m_1,n_1}+\Gamma^{(1)}_{m_2,n_2}}.
\nonumber \\ \label{eq:left2}
\end{eqnarray}
On the other hand, the \eqref{eq:right} can be written as
\begin{eqnarray}
&&
\sum_{m,n}\left\{\Omega(\Gamma^{(2)}_{m,n})\hat{e}_{\Gamma^{(2)}_{m,n}} + \frac{1}{4}\Omega(\Gamma^{(1)}_{m,n})\hat{e}_{2\Gamma^{(1)}_{m,n}} \right\} + \frac{1}{2}\left\{\sum_{n,m}\Omega(\Gamma^{(1)}_{m,n})\hat{e}_{\Gamma^{(1)}_{m,n}}\right\}^2
\nonumber \\
&&
+ \frac{1}{2} \!\!\!\!\!\!\!\!\!\!\!\!\!\!\!\!\!\!\!\!\!\!\! \sum_{{\arg Z(\Gamma^{(1)}_{m_1,n_1}) > \arg Z(\Gamma^{(1)}_{m_2,n_2})}} \!\!\!\!\!\!\!\!\!\!\!\!\!\!\!\!\!\!\!\!\!(-1)^{\left<\Gamma^{(1)}_{m_1,n_1},\Gamma^{(1)}_{m_2,n_2}\right>} \left<\Gamma^{(1)}_{m_1,n_1},\Gamma^{(1)}_{m_2,n_2}\right>\Omega(\Gamma^{(1)}_{m_1,n_1})\Omega(\Gamma^{(1)}_{m_2,n_2})\hat{e}_{\Gamma^{(1)}_{m_1,n_1} + \Gamma^{(1)}_{m_2,n_2}},
\nonumber \\ \label{eq:right2}
\end{eqnarray}
where we used the follwoing short hand notation:
\begin{eqnarray}
 \hat{e}_{\gamma} &=& U_{\gamma_{\pm,N}}^{\Omega(\gamma_{\pm,N})}\,e_{\gamma}\;U_{\gamma_{\pm,N}}^{-\Omega(\gamma_{\pm,N})}.
\end{eqnarray}
We should here note that 
\begin{eqnarray}
 \sum_{n,m}\widetilde{\Omega}(\Gamma_{m,n}^{(1)})e_{\Gamma^{(1)}_{m,n}} &=& \sum_{m,n}\Omega(\Gamma^{(1)}_{m,n})\hat{e}_{\Gamma^{(1)}_{m,n}},
\end{eqnarray}
which can also be shown by the Kontsevich-Soibelman wall-crossing formula \cite{Nishinaka:2010qk}.
Then the equivalence of \eqref{eq:left2} and \eqref{eq:right2} reduces to the desired result \eqref{eq:WCF2}.

\end{appendix}


\begin{thebibliography}{59}


\bibitem{Szendroi} B.~Szendroi, ``Non-commutative Donaldson-Thomas theory and the conifold,'' Geom.\ Topol.\  {\bf 12} (2008) 1171 [arXiv:0705.3419 [math.AG]].
\bibitem{Diaconescu-Moore}   E.~Diaconescu and G.~W.~Moore, ``Crossing the Wall: Branes vs. Bundles,'' arXiv:0706.3193 [hep-th].
\bibitem{Jafferis-Saulina} D.~L.~Jafferis and N.~Saulina, ``Fragmenting D4 branes and coupled q-deformed Yang Mills,'' arXiv:0710.0648 [hep-th].
\bibitem{Cheng-Verlinde} M.~C.~N.~Cheng and E.~P.~Verlinde, ``Wall Crossing, Discrete Attractor Flow, and Borcherds Algebra,'' SIGMA {\bf 4} (2008) 068, [arXiv:0806.2337 [hep-th]].
\bibitem{Andriyash-Moore} E.~Andriyash and G.~W.~Moore, ``Ample D4-D2-D0 Decay,'' arXiv:0806.4960 [hep-th].
\bibitem{NN} K.~Nagao and H.~Nakajima, {\it {Counting invariant of perverse coherent sheaves and its wall-crossing}},  arXiv:0809.2992[math.AG].
\bibitem{Nagao} K.~Nagao, {\it {Derived categories of small toric Calabi-Yau 3-folds and
  counting invariants}},  arXiv:0809.2994[math.AG].
\bibitem{CDWM} G.~L.~Cardoso, J.~R.~David, B.~de Wit and S.~Mahapatra, ``The mixed black hole partition function for the STU model,'' JHEP {\bf 0812} (2008) 086 [arXiv:0810.1233 [hep-th]].
\bibitem{Collinucci-Wyder} A.~Collinucci and T.~Wyder, ``The elliptic genus from split flows and Donaldson-Thomas invariants,'' JHEP {\bf 1005} (2010) 081 [arXiv:0810.4301 [hep-th]].
\bibitem{Jafferis-Moore} D.~L.~Jafferis and G.~W.~Moore, ``Wall crossing in local Calabi Yau manifolds,'' arXiv:0810.4909 [hep-th].
\bibitem{Chuang-Jafferis} W.~y.~Chuang and D.~L.~Jafferis, ``Wall Crossing of BPS States on the Conifold from Seiberg Duality and Pyramid Partitions,'' Commun.\ Math.\ Phys.\  {\bf 292} (2009) 285 [arXiv:0810.5072 [hep-th]].
\bibitem{Ooguri-Yamazaki1} H.~Ooguri and M.~Yamazaki, ``Crystal Melting and Toric Calabi-Yau Manifolds,'' Commun.\ Math.\ Phys.\  {\bf 292} (2009) 179 [arXiv:0811.2801 [hep-th]].
\bibitem{CSU} A.~Collinucci, P.~Soler and A.~M.~Uranga, ``Non-perturbative effects and wall-crossing from topological strings,'' JHEP {\bf 0911} (2009) 025 [arXiv:0904.1133 [hep-th]].
\bibitem{Dimofte-Gukov} T.~Dimofte and S.~Gukov, ``Refined, Motivic, and Quantum,'' Lett.\ Math.\ Phys.\  {\bf 91} (2010) 1 [arXiv:0904.1420 [hep-th]].
\bibitem{David} J.~R.~David, ``On walls of marginal stability in N=2 string theories,'' JHEP {\bf 0908} (2009) 054 [arXiv:0905.4115 [hep-th]].
\bibitem{Manschot1} J.~Manschot, ``Stability and duality in N=2 supergravity,'' arXiv:0906.1767 [hep-th].
\bibitem{Chuang-Pan} W.~y.~Chuang and G.~Pan, ``BPS State Counting in Local Obstructed Curves from Quiver Theory and Seiberg Duality,'' J.\ Math.\ Phys.\  {\bf 51} (2010) 052305 [arXiv:0908.0360 [hep-th]].
\bibitem{AOVY} M.~Aganagic, H.~Ooguri, C.~Vafa and M.~Yamazaki, ``Wall Crossing and M-theory,'' arXiv:0908.1194 [hep-th].
\bibitem{Herck-Wyder} W.~Van Herck and T.~Wyder, ``Black Hole Meiosis,'' JHEP {\bf 1004} (2010) 047 [arXiv:0909.0508 [hep-th]].
\bibitem{Nagao-Yamazaki} K.~Nagao and M.~Yamazaki, ``The Non-commutative Topological Vertex and Wall Crossing Phenomena,'' arXiv:0910.5479 [hep-th].
\bibitem{Sulkowski} P.~Sulkowski, ``Wall-crossing, free fermions and crystal melting,'' arXiv:0910.5485 [hep-th].
\bibitem{Lee-Yi} S.~Lee and P.~Yi, ``A Study of Wall-Crossing: Flavored Kinks in D=2 QED,'' JHEP {\bf 1003} (2010) 055 [arXiv:0911.4726 [hep-th]].
\bibitem{Aganagic-Yamazaki} M.~Aganagic and M.~Yamazaki, ``Open BPS Wall Crossing and M-theory,'' Nucl.\ Phys.\  B {\bf 834} (2010) 258 [arXiv:0911.5342 [hep-th]].
\bibitem{DGS} T.~Dimofte, S.~Gukov and Y.~Soibelman, ``Quantum Wall Crossing in N=2 Gauge Theories,'' arXiv:0912.1346 [hep-th].
\bibitem{Szabo:2009vw}
  R.~J.~Szabo, ``Instantons, Topological Strings and Enumerative Geometry,'' arXiv:0912.1509 [hep-th].
\bibitem{Krefl} D.~Krefl, ``Wall Crossing Phenomenology of Orientifolds,'' arXiv:1001.5031 [hep-th].
\bibitem{CDP1} W.~y.~Chuang, D.~E.~Diaconescu and G.~Pan, ``Rank Two ADHM Invariants and Wallcrossing,'' arXiv:1002.0579 [math.AG].
\bibitem{Manschot2} J.~Manschot, ``Wall-crossing of D4-branes using flow trees,'' arXiv:1003.1570 [hep-th].
\bibitem{CDP2} W.~y.~Chuang, D.~E.~Diaconescu and G.~Pan, ``Motivic Wallcrossing and Cohomology of The Moduli Space of Hitchin Pairs,'' arXiv:1004.4195 [math.AG].
\bibitem{Ooguri:2010yk}
  H.~Ooguri, P.~Sulkowski and M.~Yamazaki, ``Wall Crossing As Seen By Matrix Models,'' arXiv:1005.1293 [hep-th].
\bibitem{Aganagic-Schaeffer} M.~Aganagic and K.~Schaeffer, ``Wall Crossing, Quivers and Crystals,'' arXiv:1006.2113 [hep-th].
\bibitem{Nishinaka:2010qk}
  T.~Nishinaka and S.~Yamaguchi,
  ``Wall-crossing of D4-D2-D0 and flop of the conifold,''
  JHEP {\bf 1009} (2010) 026
  [arXiv:1007.2731 [hep-th]].


\bibitem{Denef:2000nb}
  F.~Denef, ``Supergravity flows and D-brane stability,'' JHEP {\bf 0008} (2000) 050 [arXiv:hep-th/0005049].
\bibitem{Denef:2002ru}
  F.~Denef,
  ``Quantum quivers and Hall / hole halos,''
  JHEP {\bf 0210 } (2002)  023.
  [hep-th/0206072].
\bibitem{Denef-Moore} F.~Denef and G.~W.~Moore, ``Split states, entropy enigmas, holes and halos,'' arXiv:hep-th/0702146.

\bibitem{AdS3-S2} J.~de Boer, F.~Denef, S.~El-Showk, I.~Messamah and D.~Van den Bleeken, ``Black hole bound states in AdS$_3$ x S$^2$,'' JHEP {\bf 0811} (2008) 050 [arXiv:0802.2257 [hep-th]].
\bibitem{BEMB} J.~de Boer, S.~El-Showk, I.~Messamah and D.~Van den Bleeken, ``Quantizing N=2 Multicenter Solutions,'' JHEP {\bf 0905} (2009) 002 [arXiv:0807.4556 [hep-th]].


\bibitem{Kontsevich-Soibelman} M.~Kontsevich and Y.~Soibelman, ``Stability structures, motivic Donaldson-Thomas invariants and cluster transformations,'' arXiv:0811.2435.
\bibitem{Kontsevich-Soibelman2} M.~Kontsevich and Y.~Soibelman, ``Motivic Donaldson-Thomas invariants: summary of results,'' arXiv:0910.4315 [math.AG].


\bibitem{GMN1} D.~Gaiotto, G.~W.~Moore and A.~Neitzke, ``Four-dimensional wall-crossing via three-dimensional field theory,'' arXiv:0807.4723 [hep-th].
\bibitem{GMN2} D.~Gaiotto, G.~W.~Moore and A.~Neitzke, ``Wall-crossing, Hitchin Systems, and the WKB Approximation,'' arXiv:0907.3987 [hep-th].
\bibitem{Cecotti-Vafa}
  S.~Cecotti and C.~Vafa, ``BPS Wall Crossing and Topological Strings,'' arXiv:0910.2615 [hep-th].
\bibitem{GMN3} D.~Gaiotto, G.~W.~Moore and A.~Neitzke, ``Framed BPS States,'' arXiv:1006.0146 [hep-th].
\bibitem{Andriyash:2010qv}
  E.~Andriyash, F.~Denef, D.~L.~Jafferis and G.~W.~Moore,
  ``Wall-crossing from supersymmetric galaxies,''
  arXiv:1008.0030 [hep-th].
\bibitem{Andriyash:2010yf}
  E.~Andriyash, F.~Denef, D.~L.~Jafferis and G.~W.~Moore,
  ``Bound state transformation walls,''
  arXiv:1008.3555 [hep-th].

\bibitem{Aspinwall:2004jr}
  P.~S.~Aspinwall,
  ``D-branes on Calabi-Yau manifolds,''
  [hep-th/0403166].


\bibitem{Gopakumar-Vafa1} R.~Gopakumar and C.~Vafa, ``M-theory and topological strings. I,'' arXiv:hep-th/9809187.
\bibitem{Gopakumar-Vafa2}  R.~Gopakumar and C.~Vafa, ``M-theory and topological strings. II,''arXiv:hep-th/9812127.



\bibitem{AOSV} M.~Aganagic, H.~Ooguri, N.~Saulina and C.~Vafa, ``Black holes, q-deformed 2d Yang-Mills, and non-perturbative topological strings,'' Nucl.\ Phys.\  B {\bf 715} (2005) 304 [arXiv:hep-th/0411280].
\bibitem{Vafa-Witten} C.~Vafa and E.~Witten, ``A Strong coupling test of S duality,'' Nucl.\ Phys.\  B {\bf 431} (1994) 3 [arXiv:hep-th/9408074].
\bibitem{Vafa1} C.~Vafa, ``Two dimensional Yang-Mills, black holes and topological strings,'' arXiv:hep-th/0406058.



\bibitem{MSW} J.~M.~Maldacena, A.~Strominger and E.~Witten, ``Black hole entropy in M-theory,'' JHEP {\bf 9712} (1997) 002 [arXiv:hep-th/9711053].



\bibitem{Nakajima}H.~Nakajima, ``Instantons on ALE spaces, quiver varieties
and Kac-Moody algebras,'' Duke\ Math.\ J.\ 76 (1994) 365.
\bibitem{Jafferis:2006ny}
 D.~Jafferis, ``Crystals and intersecting branes,'' arXiv:hep-th/0607032.




\end{thebibliography}
\end{document}